\documentstyle[apj,times,psfig]{article}

\def\simg{\mathrel{\rlap{\raise 0.511ex \hbox{$>$}}{\lower 0.511ex \hbox{$\sim$}}}}
\def\siml{\mathrel{\rlap{\raise 0.511ex \hbox{$<$}}{\lower 0.511ex \hbox{$\sim$}}}}
\def\veps{\varepsilon} \def\eps{\epsilon}
\def\tildaG{\tilde{\Gamma}} \def\const{\rm const}
 
\def\o{{\rm o}} \def\oo{{{\rm o},2}} 
\def\ergcm2s{\;\rm erg\, cm^{-2}\, s^{-1}}
 \def\phi5{\Phi_{-5}}

\begin{document}

\parskip 5pt
\topmargin 1cm

\title{"Self-absorbed" GeV light-curves of Gamma-Ray Burst afterglows}

\author{A. Panaitescu, W.T. Vestrand, P. Wo\'zniak}
\affil{ Space \& Remote Sensing, MS B244, Los Alamos National Laboratory,
       Los Alamos, NM 87545, USA}

\begin{abstract}
 We investigate the effect that the absorption of high-energy (above 100 MeV) photons produced 
 in GRB afterglow shocks has on the light-curves and spectra of Fermi-LAT afterglows.
 Afterglows produced by the interaction of a relativistic outflow with a wind-like medium 
 peak when the blast-wave deceleration sets in, and the afterglow spectrum {\sl could} be 
 hardening before that peak, as the optical thickness to pair-formation is decreasing.
 In contrast, in afterglows produced in the interaction with a homogeneous medium, the optical
 thickness to pair-formation should increase and yield a light-curve peak when it reaches unity, 
 followed by a fast light-curve decay, accompanied by a spectral softening. If energy is injected 
 in the blast-wave, then the accelerated increase of the optical thickness yields a convex afterglow 
 light-curve. Other features, such as a double-peak light-curve or a broad hump, can arise 
 from the evolution of the optical thickness to photon-photon absorption. 
 Fast decays and convex light-curves are seen in a few LAT afterglows, but the expected spectral 
 softening is rarely seen in (and difficult to measure with) LAT observations. Furthermore, for 
 the effects of photon-photon attenuation to shape the high-energy afterglow light-curve without 
 attenuating it too much, the ejecta initial Lorentz factor must be in a relatively narrow range 
 (50-200), which reduces the chance of observing those effects.
\end{abstract}


\section{Introduction}

 The first Fermi-LAT catalog of GRBs (Ackermann et al 2013a) presents temporal and spectral information
on the emission above 100 MeV measured by the Fermi satellite during the prompt (burst) and extended
(afterglow) phases of 28 GRBs. For several well-monitored sources, the high-energy emission displays
a power-law decay in time. That behaviour and the X-ray and GeV fluxes being compatible with each 
other within the blast-wave afterglow model, prompted Kumar \& Barniol Duran (2009, 2010) to attribute 
the LAT high-energy emission to the forward-shock that powers the afterglow emission at lower (optical 
and X-ray) photon energies (see also Ghirlanda et al 2010). 

 For four bright Fermi long-GRBs (080916C - Ackermann et al 2013a; 090902B - Abdo et al 2009a; 090926A - 
Ackermann et al 2011; 110731 - Ackermann et al 2013b), the LAT prompt-emission spectrum (above 100 MeV)
is a power-law harder than the Band spectrum of the GBM emission (above $\sim 300$ keV). For three of
those bursts, the slope of the LAT power-law spectrum does not change (within measurement errors) at
the end of the prompt phase. Furthermore, the LAT light-curve is continuous at the end of the burst, 
in all four cases. These two facts indicate that, at least for bright long-GRBs, the LAT emission 
mechanism is the same as for the afterglow (i.e. the forward-shock).

 The absorption of high-energy photons, discussed by Fenimore et al (1993) for CGRO-BATSE bursts, 
has been reconsidered for the GeV photons of Fermi-LAT bursts (e.g. Abdo et al 2009) and used to set 
lower limits on the {\sl burst} source Lorentz factor, usually found to be between 200 and 1000 
(e.g. Ackermann et al 2013a). A detailed treatment of the source dynamics has been shown to reduce 
those limits (Granot et al 2008, 
Hascoet et al 2012), 
as does the possibility that the burst MeV emission (the target photons for the GeV test photons) 
arose at a smaller radius than the LAT emission (Zou et al 2011, Hascoet et al 2012).

 Direct evidence for photon-photon attenuation in LAT afterglows exists only for the afterglow 090926A 
(Ackermann et al 2011), whose spectrum shows a cut-off at 2 GeV. The optical thickness to pair-formation
should vary with time, yielding a variable cut-off energy in the LAT spectrum and a variable slope of 
the attenuated spectrum. An evolving optical thickness can also alter the flux of the emergent emission, 
and the attenuated light-curve should be correlated with the spectral evolution. The purpose of this 
article is to identify the effects that pair-formation has on the afterglow light-curve, and the
correlation between features of the attenuated light-curve and the spectrum evolution.
First, we calculate the optical thickness to photon-photon absorption of the forward-shock emission
on itself, quantifying its flux with the aid of LAT measurements (taken as a proxy for the unattenuated, 
intrinsic afterglow emission). Then, we consider how that optical thickness evolves while the forward-shock 
flux, radius, and Lorentz factor vary, and study the attenuated forward-shock light-curve. Finally, we 
compare the expected features of the attenuated light-curve and the corresponding spectral evolution with 
Fermi-LAT observations of GRB afterglows.

\section{How pair-formation opacity alters afterglow light-curves}

 First, we argue that most of the target photons that absorb a test-photon (of energy $\veps$ 
above 100 MeV) were produced within the last dynamical timescale since the test-photon was released. 
The kinematics of the interaction between a target-photon emitted at radius $R_\o$ by a spherical surface,  
moving at Lorentz factor $\Gamma (R)$, interacting with a test-photon at radius $R$ shows that the angle 
(relative to the radial direction) at which the target-photon was emitted is
\begin{equation}
  \gamma = \Gamma_\o^{-1} \left\{ \begin{array}{ll} 
           (R/R_\o)^{1/2}     & \Gamma = \const \\
           (1/\sqrt{2}) (R/R_\o)  & \Gamma \propto r^{-1/2} \\
           (1/2) (R/R_\o)^2  & \Gamma \propto r^{-3/2}  \\
                \end{array} \right.
\label{gamma}
\end{equation}
with $\Gamma_\o = \Gamma (R_\o)$. The first line is for an undecelerated source, the second line is 
for a decelerating shock interacting with a wind-like ambient medium (with density radial distribution 
$\rho \propto r^{-2}$), while the third line is for a decelerating shock interacting with a homogeneous 
medium. Furthermore, the angle of incidence between the target and test photons is $\delta = \gamma (R_\o/R)$ 
and the target-photon was emitted from a region extending an angle $\theta = \gamma - \delta$, as seen
from the origin.

 The infinitesimal optical thickness to pair-formation for a test-photon of energy $\eps$ crossing 
a distance $dr$ of the target-photon front is $d\tau \propto n(>\eps_t)|_R (1-\cos \delta) dr$, where
$n(>\eps_t)|_R$ is the density of target photons above the threshold energy for pair-formation,
$\eps_t \propto [\eps (1 -\cos \delta)]^{-1}$, proportional to the flux of target photons at radius $R$.
Denoting by $D_\o = [\Gamma_\o (1 - v_\o \cos \gamma)]^{-1} \propto (\Gamma_\o \gamma^2)^{-1}$ the 
relativistic boost factor, we have $n(>\eps_t)|_R \propto D_\o^3 I'(>\eps_t/D_\o) 
\Delta \Omega$, where $I'(>\eps_t/D_\o)$ is the comoving-frame intensity of the target photons 
above the threshold energy and $\Delta \Omega \simeq (R_\o \theta/R)^2$ is the solid-angle 
extended at $R$ by the source of target photons (at $R_\o$). If the target photons have a
power-law number spectrum above an energy $\eps'_p$, $I'(\eps') = I'_p (\eps'/\eps'_p)^{-(\beta+1)}$,
then $I'(>\eps_t/D_\o) \simeq I'_p (\eps'_p)^{\beta+1} (D_\o/\eps_t)^\beta$. For a small incidence angle,
$\delta \sim \Gamma_\o^{-1} \ll 1$, we have $\theta \simeq \gamma$ and
\begin{equation}
 \frac{d\tau}{dr} \propto I'_p (\eps'_p)^{\beta+1} \frac{(R_\o/R)^{2\beta+4}}{\Gamma_\o^{\beta+3}\gamma^2}
\label{cic}
\end{equation}

 To calculate the optical thickness to pair-formation, we must evaluate the term 
$I'_p (\eps'_p)^{\beta+1}$ at $R_\o$. For synchrotron emission, $I'_p \eps'_p \propto
B N_e$ and $\eps'_p \propto B E^2$, where $B$ is the magnetic field, $N_e$ is the number of
radiating electrons, and $E$ is the comoving-frame energy of electrons radiating synchrotron
emission at energy $\eps'_p$. If the magnetic field and the electrons acquire a constant 
fraction of the forward-shock's energy, then $B \propto \Gamma_\o \sqrt{\rho(R_\o)}$ and 
$E \propto \Gamma_\o$. The number of radiating electrons is $N_e \propto R_\o^3 \rho(R_\o)$.

 With the above scalings, it can be shown that
\begin{equation} \begin{array}{ll} 
  I'_p \eps'_p = \const \;,\;  \eps'_p \propto R_\o^{-1} & (\Gamma = \const, \rho \propto r^{-2}) \\
  I'_p \eps'_p \propto R_\o^3 \;,\;  \eps'_p = \const         & (\Gamma = \const, \rho \propto r^0) \\
  I'_p \eps'_p \propto \Gamma_\o \;,\; \eps'_p \propto \Gamma_\o^3 R_\o^{-1} & (\Gamma \propto r^{-1/2}, 
                                                                            \rho \propto r^{-2}) \\
  I'_p \eps'_p \propto \Gamma_\o R_\o^3 \;,\; \eps'_p \propto \Gamma_\o^3 & (\Gamma \propto r^{-3/2}, 
                                                                          \rho \propto r^0) \\
                \end{array} 
\end{equation}
Substituting in equation (\ref{cic}), we find  
\begin{equation}
  \frac{d\tau}{dr} \propto \left\{ \begin{array}{ll} 
           R_\o^{\beta+5}  & (\Gamma = \const, \rho \propto r^{-2})    \\
           R_\o^{2\beta+8} & (\Gamma = \const, \rho  \propto r^0)    \\
           R_\o^6          & (\Gamma \propto r^{-1/2}, \rho \propto r^{-2} ) \\
           R_\o^{11-\beta} & (\Gamma \propto r^{-3/2}, \rho  \propto r^0 ) \\ 
                \end{array} \right.
\end{equation}
Thus, $d\tau/dr$ is a strong function of $R_\o$, which shows that most of the target photons are
those emitted at $R_\o \siml R$. Their typical incidence angle on the test-photon is $\delta \simeq \Gamma^{-1}$.

\subsection{Approximate optical thickness to pair-formation}

 The optical thickness to pair-formation for a test-photon of {\sl observer}-frame energy $\veps$ is
$\tau (\veps)= \overline{\sigma} N [>\eps_t(\veps)]$, with $\overline{\sigma}$ the photon-photon
absorption cross-section averaged over the power-law spectrum of the target photons with energy above 
the {\sl source}-frame (at redshift $z$) threshold $\eps_t$ and $N$ the fluence of target photons at 
the location $R$ where the test-photon is absorbed.

 The threshold energy for pair-formation is
\begin{equation}
 \eps_t (\veps) = \frac{2 (m_e c^2)^2}{(z+1) \veps (1-\cos \delta)} = \frac{4 \Gamma^2 (m_e c^2)^2}{(z+1) \veps} 
\end{equation}
where we used $\delta \simeq \Gamma^{-1} \ll 1$.
In the observer-frame, the threshold photon energy 
\begin{equation}
 \veps_t = \frac{\eps_t}{z+1} = 12 \left( \frac{z+1}{3} \right)^{-2} \left( \frac{\Gamma}{100} \right)^2 
                      \left( \frac{\veps}{\rm{100\,MeV} }\right)^{-1} \; {\rm MeV}
\label{epst}
\end{equation}
is not far below 100 MeV, if the source Lorentz factor $\Gamma$ is around 100, thus it is reasonable to 
assume that the spectrum of target photons is the extrapolation to lower energies of the afterglow 
spectrum measured by LAT above 100 MeV. Then, the fluence of target photons above the threshold energy is
\begin{equation}
 N(>\eps_t) = \left( \frac{\eps_t}{(z+1)\veps} \right)^{-\beta} N[>(z+1)\veps] 
\end{equation}

 Considering that the target photons are those produced within the last dynamical timescale,
the above number of photons incident per unit area at the location $R$ of a test-photon is 
\begin{equation}
 N[>(z+1)\veps] \simeq \frac{1}{4\pi R^2} \frac{L[>(z+1)\veps]}{(z+1)\veps} \frac{t}{z+1}
\end{equation}
for spectra with $\beta > 1$, and with $t$ the observer-frame time. The luminosity $L$ can be 
calculated from the measured flux: 
$L(>(z+1)\veps) \simeq 4\pi d_l^2 F_\o(> \veps)$, where $F_\o(> \veps)$ is the afterglow {\sl unabsorbed} 
energy flux and $d_l$ the luminosity distance.

 Finally, the optical thickness to pair-formation for a photon of measured energy $\veps$ can be 
related to the observed afterglow flux above energy $\veps$: 
\begin{equation}
  \tau (\veps) = \overline{\sigma} (z+1)^{2(\beta -1)} \left( \frac{d_l}{R} \right)^2 
         \frac{t F_\o(>\veps)}{(2 \Gamma m_e c^2)^{2\beta}}\; \veps^{2\beta-1} 
\label{tau0}
\end{equation}
To calculate the optical thickness from the flux $F_\o$ measured above a photon energy $\veps_0 \neq \veps$, 
one can substitute $F_\o(>\veps) = F_\o(>\veps_0) (\veps/\veps_0)^{1-\beta}$ in equation (\ref{tau0}),
leading to 
\begin{equation}
 \tau (\veps, t) \propto \frac{t F_\o(> \veps_0)}{\Gamma^{2\beta}}\; \veps_0^{\beta -1} \veps^{\beta} \;.
\label{final}
\end{equation}

 We simplify equation (\ref{tau0}) by considering $\beta = 1$ (compatible with the slope measured by 
LAT for many afterglows),
for which $\overline{\sigma} = 0.18\, \sigma_e = 1.2 \times 10^{-25}\; {\rm cm^2}$, and by using 
$d_l = 5 \times 10^{27} (z+1)^2$ cm (this approximation has an error lower than 25 percent for
$0.5 < z < 5$): 
\begin{equation} 
 \tau(\veps)  \simeq 0.15 \left( \frac{z+1}{3} \right)^4 \frac{F_{-6} t_1}{R_{16}^2 \Gamma_2^2} \veps_8 \\
\label{tau}
\end{equation}
with the notations $X_n = 10^{-n} X({\rm cgs})$ and $\veps_8 = \veps/({\rm 100 MeV})$. 
The afterglow unabsorbed energy flux above the test-photon energy $\veps$ is normalized to 
$10^{-6} \ergcm2s$, which, for $\beta = 1$, corresponds to a photon flux
$C (>\veps) = F_\o(> \veps)/(4.6\, \veps) \simeq 10^{-3} \ergcm2s$ in the 100 MeV--10 GeV range, 
as typically measured by LAT at $\sim$ 10 s after trigger.

 Given some of the approximations made, the coefficient given in equation (\ref{tau}) is likely
mis-estimated by a factor of about 2, but that leads to a small error in the following results
for the ejecta Lorentz factor, because that coefficient is raised to a small, sub-unity power. 
A larger error is made if the afterglow power-law spectrum measured by LAT above 100 MeV does 
not extend sufficiently below 100 MeV, which may happen if the LAT emission process is inverse-Compton.
Alternatively, if the LAT emission is synchrotron, then afterglow X-ray observations (which indicate 
that the peak of synchrotron spectrum is below 1 keV at 100--1000 s) imply that the synchrotron
power-law spectrum is well below 100 MeV even at the earliest LAT measurement.

\subsection{Burst vs. afterglow target photons}

 Equation (\ref{tau0}) quantifies the optical thickness to pair-formation for a LAT test-photon absorbed 
by a target-photon that arose from the same mechanism as the LAT emission, i.e. a target-photon that is
the extrapolation of the LAT spectrum below 100 MeV (equation \ref{epst}).

{\sl During the burst}, the absorption of (more than) 100 MeV LAT photons on (less than) 10 MeV GRB photons 
can be estimated from equation 
(\ref{final}): 
\begin{equation}
 \frac{\tau_{grb}(\veps)}{\tau_{lat}(\veps)} = \frac{ \Phi_{grb} }{ \Phi_{lat} } 
             \left(  \frac{ \Gamma_{aglow}^{\beta_{lat}} }{ \Gamma_{grb}^{\beta_{grb}} } \right)^2
         \frac{\veps_p^{\beta_{grb}-1}}{\veps_o^{\beta_{lat} -1}} \veps^{\beta_{grb} - \beta_{lat}}
\label{grbaglow}
\end{equation}
where $\Phi$ denotes the fluences of each component, $\veps_p$ is the peak energy of the prompt emission,
$\veps_\o$ is the low edge of the LAT range, all photon energies are in $m_e c^2$ units, and we allowed for 
different Lorentz factors of the prompt and afterglow sources.

 The LAT afterglow fluence $\Phi_{lat}$ during the prompt phase is 1--50 percent of the 10 keV--1 MeV 
prompt emission fluence $\Phi_{grb}$ (Ackermann et al 2013a).
The GRB spectrum being softer ($\beta_{grb} \simeq 2$) than the spectrum of the LAT afterglow 
($\beta_{lat} \simeq 1$), decreases the relative number of burst photons above the threshold of $\sim 10$ MeV
(equation \ref{epst}) for pair-formation by a factor $(\Gamma_{aglow}^2/\Gamma_{grb}^4)\, \veps_p \veps \simeq 
\veps/\Gamma^2 = 0.01\, \veps_8 (\Gamma/100)^{-2}$, for $\veps_p = 1$ and $\Gamma_{aglow} \simeq \Gamma_{grb} 
\equiv \Gamma$. 

 Given that the afterglow source is unlikely to be more relativistic than the GRB source ($\Gamma_{aglow}
\siml \Gamma_{grb}$), it follows that, for the brighter LAT afterglows ($\Phi_{lat} \simg 0.1 \Phi_{grb}$), 
pair-formation opacity for 100 MeV photons on burst photons is negligible relative to that on afterglow 
photons, provided that the LAT power-law spectrum extends below $\sim 10$ MeV (equation \ref{epst}).
However, for the dimmer LAT afterglows ($\Phi_{lat} < 0.01 \Phi_{grb}$), pair-formation on burst photons
should be dominant. In that case, the lower limits on the afterglow source initial Lorentz factor $\Gamma_\o$
derived below (from the condition of optical thinness for the 100 MeV afterglow photons) are
underestimations. 

 {\sl After the burst}, a LAT test-photon can be absorbed by a GRB target-photon, even though the
former was emitted (at $R$) after the latter (at $R_o < R$), provided that GRB photon moves at a sufficiently 
large angle $\gamma$ relative to the LAT photon's direction (toward the observer). According to equation
(\ref{gamma}), the GRB photon's emission angle must be larger than the inverse of the GRB source Lorentz
factor, which implies that the burst emission absorbing the LAT afterglow emission is highly "de-beamed" 
relativistically, by a factor $D_o^3 \propto (\Gamma_{grb} \gamma)^{-6} \ll 1$. Adding that to equation 
(\ref{grbaglow}) and taking into account that the afterglow fluence $\Phi_{lat}$ is only slowly decreasing 
after the burst, it follows that the absorption of LAT photons by the GRB emission can be ignored safely.

\subsection{Forward-shock emission}

 The interaction of the relativistic ejecta with the ambient medium drives a reverse shock into the ejecta
shell and forward shock that sweeps the ambient medium. Before the reverse shock sweeps up the ejecta
shell, both shocked fluids move at the same and constant Lorentz factor. After the reverse shock has
crossed the ejecta shell, the motion of the shocked fluids is significantly decelerated by the interaction 
with ambient medium. The deceleration timescale $t_d$, defined by the reverse shock having crossed the ejecta
shell, is also the time when a substantial fraction of the ejecta energy has been transferred to the
swept-up ambient medium. 
 For simplicity, we consider only the case of a sufficiently thin/dense shell of relativistic ejecta;
then, at $t < t_d$, the shocked fluid moves at nearly the Lorentz factor $\Gamma_\o$ of the unshocked
ejecta. In the case of a thick/tenuous ejecta shell, the pre-deceleration Lorentz factor of the shocked 
fluid depends on the geometrical thickness $\Delta$ of the ejecta shell and the following calculation 
would yield a constraint on $\Delta$ instead of $\Gamma_\o$.

 To relate the observer arrival-time of the forward-shock photons to that shock's radius, 
we note that most photons are emitted from the edge of the "visible" area, where the radiating fluid 
moves at an angle $\theta = \Gamma^{-1}$ relative to the direction toward the observer. The 
electrons radiating synchrotron emission above 100 MeV are most likely cooling on timescale smaller 
than the dynamical time, hence they are located immediately behind the forward-shock, which moves that 
$\Gamma_{fs} = \sqrt{2} \Gamma$. Then, at $t < t_d$, when the source moves at constant $\Gamma_\o$,
the photon arrival-time $ct = (z+1) (R/2) (\Gamma_{fs}^{-2} + \theta^2)$ is
\begin{equation}
  t = \frac{3}{4} (z+1) \frac{R}{c\Gamma_\o^2}
\label{time}
\end{equation}
The shock radius $R$ can be substituted in equation (\ref{tau}), leading to
\begin{equation}
 \tau (t<t_d) \simeq  0.1 \left( \frac{z+1}{\Gamma_\oo} \right)^6 \frac{F_{-6}}{t_0} \veps_8 \;.
\label{ttd}
\end{equation}

 To continue, the time-evolution of $F_\o$ is needed. For that, we assume that the afterglow emission 
above 100 MeV is synchrotron from the forward-shock that energizes the ambient medium, but the 
calculations below for the attenuated flux can be easily extended to the inverse-Compton emission
or to the reverse shock emission, as long as the target-photon (below 10 MeV) and the test-photon 
(above 100 MeV) are from the same spectral component (i.e. they arise from same emission 
mechanism and same radiative process). This condition allows us to relate the flux of absorbing 
photons (of energy below the LAT range) to that measured by LAT.

\begin{figure*}[t]
\centerline{\psfig{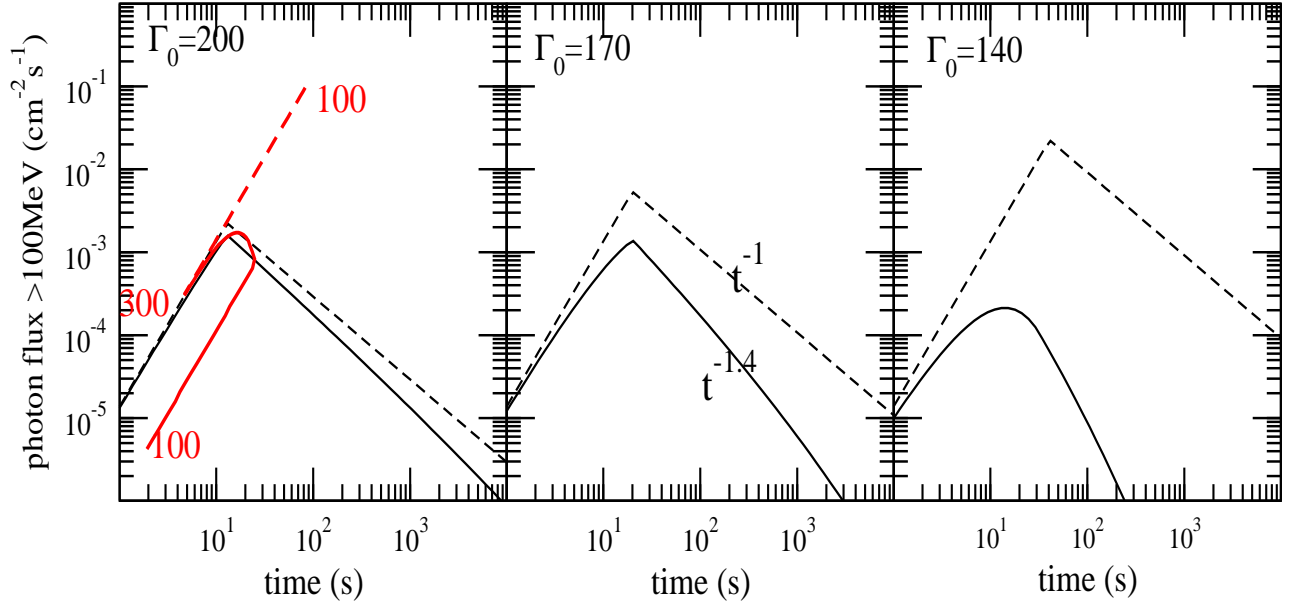}}
\figcaption{ Effect of photon-photon attenuation on the 100 MeV--10 GeV synchrotron emission
   produced by the forward shock interacting with a homogeneous medium. Dashed curves show
   the intrinsic afterglow flux, solid lines are for the attenuated flux. The synchrotron
   cooling frequency is assumed to be below the LAT range. The blast-wave parameter is
   $E/n = 10^{50}\; {\rm erg \, cm^{-3}}$ (and $z=2$).
   The intrinsic forward-shock emission rises at $t^2$ until $t_d$, when it peaks, and decays
   as $t^{-1}$ (for $\beta = 1$) after that, with the photon flux being fixed at
   $10^{-3} \, {\rm cm^{-2}\, s^{-1}}$ at 10 s.
   As the ejecta initial Lorentz factor $\Gamma_\o$ decreases (from left to right),
   pair-formation has a stronger effect, yielding attenuated light-curves that are dimmer and
   peak earlier. That is summarized by the red curves in the left panel, showing the peak flux
   vs peak time for $\Gamma_\o$ ranging from 100 to 300 (the upper limit on the peak time,
   given in equation \ref{tpmax}, is $t_p^{(max)} \simeq 20$ s).
   The optical thickness to pair-formation $\tau$ (the vertical separation between the
   intrinsic and attenuated light-curves is the absorption factor $e^\tau$) increases with time. 
   For a higher $\Gamma_\o$ (left panel),
   $\tau$ exceeds unity only after $t_d$, and yields a steeper decay for the attenuated flux.
   For a smaller $\Gamma_\o$ (right panel), $\tau$ reaches unity before $t_d$ and yields the peak
   of the attenuated light-curve. }
\end{figure*}

\subsection{Homogeneous medium}

 Prior to the onset of deceleration and at photon energies above the cooling frequency of the 
synchrotron spectrum, the forward-shock synchrotron flux should rise as $F_\o \propto t^2$.
Below the cooling frequency, the rise is $t^3$; the following calculations would be very similar, 
and all the important features of the photon-photon attenuation on the emergent light-curve
can be captured by the low cooling-frequency case.
Thus, if we parametrize the flux above 100 MeV by $F_\o (t < t_d)= 10^{-6} F_{-6} (t/10)^2 \ergcm2s$, 
to match LAT measurements at 10 s, equation (\ref{ttd}) gives 
\begin{equation}
 \tau (t<t_d) \simeq 10^{-2} (z+1)^6 \frac{F_{-6} t_1}{\Gamma_\oo^6}\, \veps_8 \;.
\label{t1}
\end{equation}

  At the deceleration radius, about 1/3 of the initial ejecta energy has been transferred to the 
energized ambient medium, whose comoving frame internal energy is larger by $\Gamma_\o$ than its 
rest-mass (this is one of the shock-jump conditions). Thus, the forward-shock's energy at $t_d$ is 
\begin{equation}
  E_{fs} = \frac{4\pi}{3} n m_p c^2 R_d^3 \Gamma_\o^2
\label{E}
\end{equation}
Substituting $R_d$ in equation (\ref{time}), the deceleration timescale is
\begin{equation}
  t_d = 830\; \frac{z+1}{3} \left(\frac{E_{53}}{n_0}\right)^{1/3} 
            \left(\frac{\Gamma_\o}{100}\right)^{-8/3} \; {\rm s} 
\label{td}
\end{equation}
After $t_d$, the afterglow intrinsic flux should decrease as $F_\o \propto t^{-\alpha}$ with
$\alpha = (3\beta-1)/2$, where $\beta$ is the slope of the energy spectral flux $F_\nu$, thus 
$\alpha = 1$ is expected for $\beta = 1$. Equation (\ref{E}) with $R$ and $\Gamma(R)$ instead of 
$R_d$ and $\Gamma_\o$, and the general result $t \propto R/\Gamma^2$ lead to $R \propto t^{1/4}$
and $\Gamma \propto t^{-3/8}$.  Then, equation (\ref{tau0}) yields $\tau (t>t_d) \propto t^{1-3\beta/4}$, 
i.e. $\tau$ increases for $\beta < 4/3$ and decreases for $\beta > 4/3$. LAT afterglows have 
$\beta \simg 1$, thus $\tau$ should increase slowly as $\tau(t>t_d) \propto t^{1/4}$ or be nearly 
constant at the value reached at $t_d$. 

 Adding to this that $\tau (t<t_d)$ increases with time, it leads to the conclusion that optical 
thickness to pair-formation plays a role in shaping the afterglow light-curve if $\tau (t_d) > 1$. 
Substituting equation (\ref{td}) in (\ref{t1}),
\begin{equation}
 \tau (t_d) = 700 \left(\frac{z+1}{3}\right)^7 \left( \frac{E_{53}}{n_0} \right)^{1/3} 
                (F_{-6}\veps_8)\, \Gamma_\oo^{-26/3} 
\end{equation}
thus, for $\Gamma_\o$ equal to
\begin{equation}
 \tildaG = 210 \left(\frac{z+1}{3}\right)^{21/26} \left( \frac{E_{53}}{n_0} \right)^{1/26} 
    (F_{-6} \veps_8)^{3/26}
\label{Gtilde}
\end{equation}
we have $\tau (t_d) = 1$.
Note that the critical value $\tildaG$ is quite well determined (if redshift is known), as other 
quantities appear at very small powers.

 If $\Gamma_\o > \tildaG$, then $\tau (t_d) < 1$, and pair-formation opacity is negligible, except 
when $\tau (t_d) \siml 1$ and could exceed it after $t_d$, owing to a slow $\tau \propto t^{1/4}$ increase 
(for $\beta = 1$). In this case, the peak of the LAT light-curve occurs at the deceleration time 
($t_p = t_d)$, and equation (\ref{td}) gives 
\begin{equation}
 \Gamma_\o (t_p = t_d) = 1250 \left(\frac{z+1}{3}\right)^{3/8} \left( \frac{E_{53}}{n_0} \right)^{1/8} 
     t_{p,0}^{-3/8}
\label{Gpk1}
\end{equation}

\begin{figure*}[t]
\parbox[h]{88mm}{ \psfig{figure=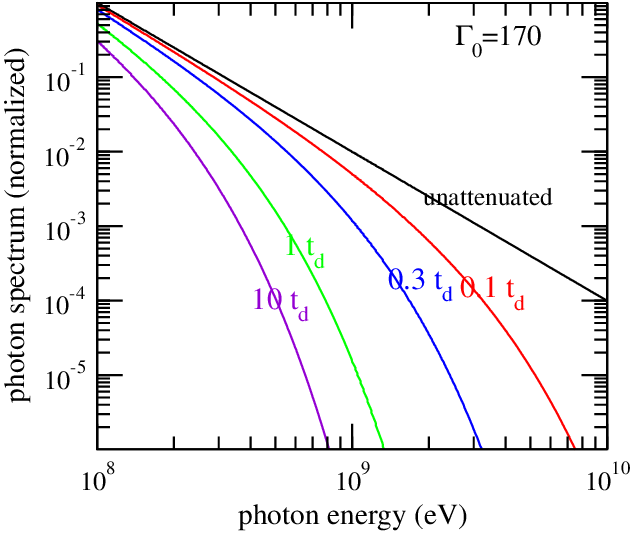,height=70mm}
\figcaption{ Spectral softening of the forward-shock synchrotron emission whose light-curve is
    shown in middle panel of Figure 1, for $\Gamma_\o =170$. The optical thickness
    to pair-formation increases with photon energy at fixed epoch (given for each curve)
    and increases in time for a fixed photon energy. The intrinsic photon spectrum has slope 2, 
    the 0.1-1 GeV attenuated spectrum has an effective slope $\simeq 3.5$ at $t_d$ 
    (the peak of the light-curve). Thus, a substantial effect of pair-opacity on the afterglow 
    flux above 100 MeV (here, an attenuated flux decay index larger by $\delta \alpha = 0.4$ 
    than the intrinsic flux) should be accompanied by a substantial spectral softening and 
   by the lack of photons above 1 GeV.  }}
\hspace*{7mm}
\parbox[h]{88mm}{ \vspace*{-17mm} \psfig{figure=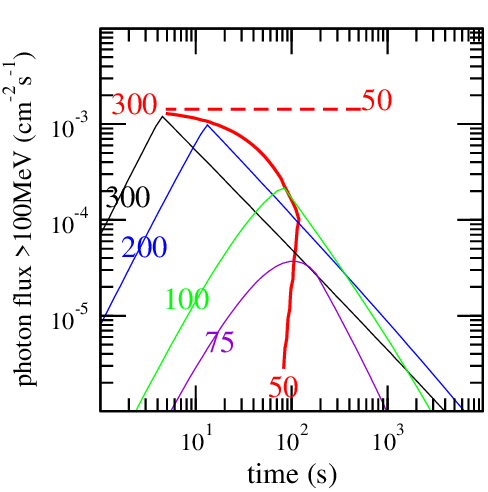,width=88mm,height=75mm}
\figcaption{ LAT light-curves as for Figure 1 but for the intrinsic afterglow flux fixed
    at $10^{-3} \, {\rm cm^{-2}\, s^{-1}}$ at its peak ($t_d$). Thick, red lines indicate the
    peak flux vs peak epoch for the attenuated flux (solid line) and intrinsic flux (dotted line)
    and for $\Gamma_\o$ ranging from 50 to 300. Thin lines are the attenuated light-curves
    for a few values of $\Gamma_\o$.  }}
\end{figure*}

 The interesting situation occurs for $\Gamma_\o < \tildaG$, for which $\tau (t_d) > 1$. 
In this case, $\tau (t<t_d)$ reaches unity at 
\begin{equation}
 t_\pm \simeq 1.2\, \left(\frac{z+1}{3}\right)^{-6} (F_{-6}\veps_8)^{-1} 
             \left(\frac{\Gamma_\o}{100} \right)^6 \;{\rm s}
\end{equation}
according to equation (\ref{t1}). After that, $\tau (t_\pm < t < t_d) = t/t_\pm$ and the absorbed 
afterglow flux evolves as $F (t) = F_\o e^{-\tau} \propto t^2 \exp (-t/t_\pm)$, 
reaching a maximum at $t_p = 2 t_\pm$, hence the timing of the attenuated light-curve peak gives 
\begin{equation}
 \Gamma_\o (t_p = 2 t_\pm) = 86\; \frac{z+1}{3} (F_{-6} \veps_8)^{1/6} t_{p,0}^{1/6} \;.
\label{Gpk2}
\end{equation}
If $\Gamma_\o < \tildaG$ and the LAT light-curve peak is erroneously identified with the deceleration
timescale, then $\Gamma_\o$ is overestimated by a factor $13\; t_p^{-13/24} [(z+1)/3]^{-5/8} 
(F_{-6} \veps_8)^{1/6} (E_{53}/n_0)^{1/8}$ (from eqs \ref{Gpk1} and \ref{Gpk2}). 

 Given that $t_p \simeq t_\pm < t_d$ for $\Gamma_\o < \tildaG$ and $t_p = t_d$ for $\Gamma_\o > \tildaG$,  
it follows that there is an upper limit to the LAT light-curve peak epoch, reached when $t_\pm =t_d$
(i.e. for $\Gamma = \tildaG$) : 
\begin{equation}
 t_p^{(max)} = 110\, \left( \frac{z+1}{3} \right)^{-15/13} \left( \frac{E_{53}}{n_0} \right)^{9/39} 
              (F_{-6} \veps_8)^{-14/13} \; {\rm s}
\label{tpmax}
\end{equation}
A somewhat lower $t_p^{(max)} \simeq 63\, [(z+1)/3]^{-15/17}...$ s (with remaining quantities above at
a small power) is obtained if the synchrotron cooling frequency were above 100 MeV. A similar
result should be obtained if the LAT emission were inverse-Compton, as that changes only the
rise of the afterglow flux before $t_d$, affecting $\tildaG$ and $t_\pm$. The existence
of an upper limit to the LAT light-curve peak epoch is a consequence of the intrinsic afterglow 
light-curve having a peak and of the optical thickness to pair-formation increasing with time 
(for a wind-like medium, that optical thickness is, most likely, decreasing in time, as discussed 
in the next section).

 Figure 1 illustrates the effect of photon-photon absorption on the emergent flux
above 100 MeV, calculated by integrating spectral flux $dF/d\veps$ attenuated by the optical 
thickness $\tau (\veps)$. For $\beta = 1$, the optical thickness increases after the deceleration 
time (the peak of the intrinsic light-curve) as $\tau \propto t^{1/4}$ and the attenuated flux 
displays a faster decay than the intrinsic light-curve. Note that, for such a slowly varying 
optical thickness, the attenuated light-curve decay, $F (t>t_d) = F_\o e^{-\tau} \propto t^{-\alpha} 
\exp\{-(t/t_d)^{1/4}\}$, there is little curvature in the attenuated light-curve, and it
still resembles a power-law. Thus, LAT light-curves decaying faster than expected for the 
forward-shock model (i.e. faster than the X-ray emission at same time), may be due photon-photon 
attenuation in the source. 

 Figure 2 shows the obvious correlation between a significant effect of photon-photon 
attenuation on the LAT light-curve (Fig 1) and the lack of higher energy photons.
Furthermore, Figure 2 illustrates the progressive softening of the afterglow spectrum 
above 100 MeV, owing to an increasing $\tau (\veps)$. Together with the intrinsic and attenuated 
light-curves shown in Figure 1, this suggests the less trivial, but still intuitive,
conclusion that LAT light-curves decaying faster should have LAT spectra that soften faster.

 Figure 3 shows the peak flux--peak epoch curve obtained by varying the ejecta initial 
Lorentz factor, but with the intrinsic afterglow flux fixed at the deceleration time (instead of 10 s). 
An upper limit to the peak epoch of the emergent light-curve still exists, being characterized
by $\tildaG \simeq 100$ and $t_p^{(max)} \simeq 100$ s. Again, for $\Gamma_\o < \tildaG$, the peak 
flux of the emergent emission decreases fast with decreasing $\Gamma_\o$.  

\begin{figure*}[t]
\parbox[h]{88mm}{ \psfig{figure=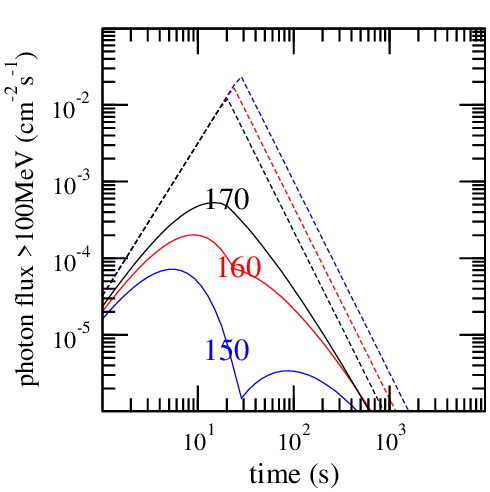,width=88mm,height=75mm}
\figcaption{ LAT light-curves as for Figure 1 but for a $\beta = 2$ spectrum. 
    The decrease of optical thickness after $t_d$ (the peak of the intrinsic light-curve -- 
    dashed lines) may yield a a second, dimmer peak in the attenuated flux light-curve (solid lines). 
    Light-curves are labelled by their $\Gamma_\o$. } }
\hspace*{7mm}
\parbox[h]{88mm}{ \vspace*{6mm} \psfig{figure=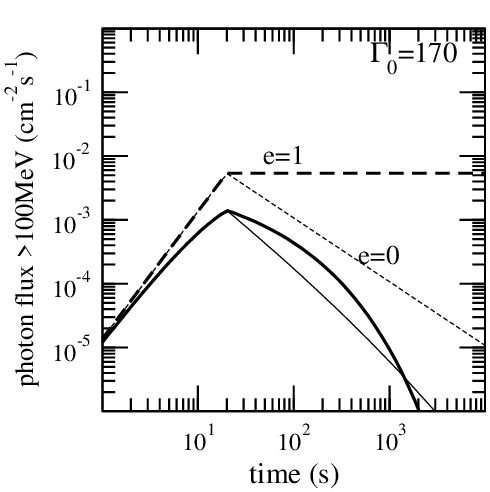,width=88mm,height=75mm}
\figcaption{ LAT light-curves as for middle panel of Figure 1, with the forward-shock 
  energy evolving as $E \propto t^e$ after $t_d$ (thin lines -- without energy injection, 
  thick lines for $e=1$). Dashed lines indicate the intrinsic afterglow emission, solid lines 
  are for the emergent flux. The evolution of the optical thickness to pair-formation
  is $\tau (t > t_d) \propto t^{(e+1)/4}$, hence $e=1$ yields a faster increase of $\tau$,
  leading to a convex attenuated light-curve.  }}
\end{figure*}

 Given that $\tau (t>t_d) \propto t^{-(3\beta/4-1)}$, the optical thickness $\tau$ decreases 
after the peak of the intrinsic light-curve if the energy spectrum slope is $\beta > 4/3$. 
The change in the evolution of $\tau$ across $t_d$ yields a point of inflection in the
attenuated light-curve if $\tau(t_d) > 1$, as illustrated in Figure 4.
For $\tau (t_d)$ not much above unity, the emergent light-curve displays a power-law decay that 
is slower than that of the intrinsic flux, but for  $\tau(t_d) \gg 1$, the attenuated light-curve
has a second peak that may be detected by LAT.
 
 Figure 5 shows how a faster evolving $\tau$ induces an evident curvature in the
attenuated light-curve. Considering that $\tau \propto t^{1-3\beta/4}$, a spectral slope $\beta$ 
sufficiently far from 4/3 should illustrate that effect. However, a faster evolving $\tau$ is 
obtained for the typical $\beta \simeq 1$ LAT spectrum, if there is an energy injection\footnotemark
in the forward shock, so that its energy evolves as $E \propto t^e$ after $t_d$.
\footnotetext{The existence of an energy injection process in the blast-wave is suggested by 
the light-curve plateaus (slow decays) observed by XRT in a good fraction of X-ray afterglows}
This energy-injection law yields a decay index of the intrinsic flux $\alpha = [3\beta -1-(\beta+1)e]/2$, 
the dynamics of the forward-shock becomes $R \propto t^{(e+1)/4}$, $\Gamma \propto t^{-(3-e)/8}$, 
and equation (\ref{tau0}) gives $\tau (t>t_d) \propto t^{1-\beta(3-e)/4}$. For $\beta =1$, $\alpha = 1-e$ 
and $\tau \propto t^{(e+1)/4}$. As shown in Figure 5, afterglows for which $\tau$ increases 
sufficiently fast after $t_d$ may disappear at late times, their flux falling below the extrapolation 
of the $F \propto t^{-\alpha}$ decay seen just after the peak.

\subsection{Wind-like medium}

 Before deceleration, the intrinsic synchrotron flux evolves as $F_\o \propto t^{-(\beta-1)}$ above 
the synchrotron cooling frequency, and as $t^{-\beta}$ below that frequency. Considering only
the former case (calculations and results are similar for the latter), and taking $\beta =1$, 
we parametrize the afterglow flux as $F_\o (t< t_d)= 10^{-6} F_{-6} t^0 \ergcm2s$.
From equation (\ref{ttd})
\begin{equation}
 \tau (t<t_d) \simeq 10^{-2} (z+1)^6 \frac{F_{-6} \veps_8 }{\Gamma_\oo t_1}
\label{t2}
\end{equation}
Thus, $\tau$ decreases with time before deceleration, a behavior which is maintained after $t_d$, 
and is enhanced if there is energy injection. This is the only important difference relative to 
the case of a homogeneous medium.

\begin{figure*}
\centerline{\psfig{figure=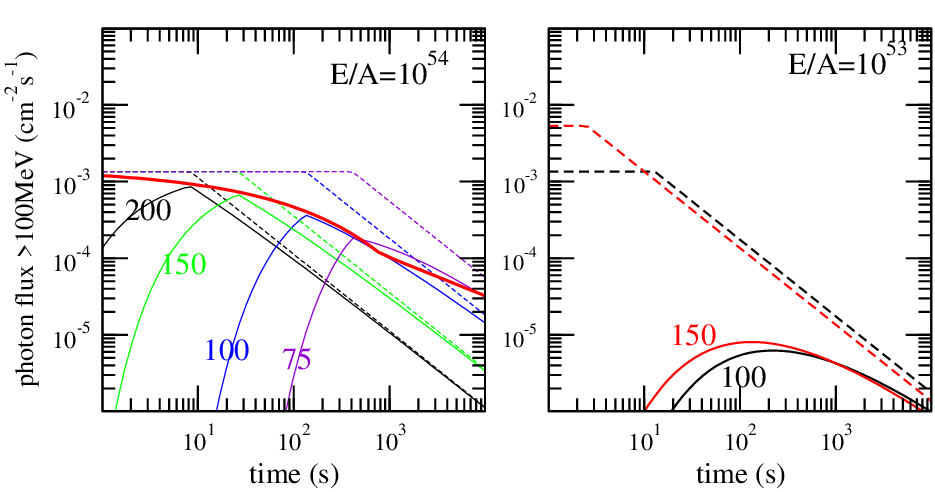,width=160mm}}
\figcaption{ Effect of photon-photon attenuation on the 100 MeV--10 GeV synchrotron emission
   produced by the forward shock interacting with a wind-like medium. Dashed curves show
   the intrinsic afterglow flux, solid lines are for the attenuated flux. The synchrotron
   cooling frequency is assumed to be below the LAT range. The blast-wave parameter is
   shown in each panel. The intrinsic forward-shock emission is flat until $t_d$, and 
   decays as $t^{-1}$ after that. The optical thickness to pair-formation decreases with time.
   Left panel: $\Gamma_\o > \tildaG \simeq 80$ and $\tau (t_d) < 1$, yielding a rise of the 
   attenuated flux. Red, thick line shows the light-curve peak flux vs peak epoch for various
   $\Gamma_\o$.
   Right panel: $\Gamma_\o < \tildaG \simeq 250$ and $\tau (t_d) > 1$, yielding a light-curve hump. 
   Light-curve are labelled by the ejecta initial Lorentz factor. }
\end{figure*}

 The deceleration timescale can be found from equations (\ref{time}) and (\ref{E}) modified 
for a wind-like medium: $E = 4\pi (3\times 10^{35}\, A) m_p c^2 R_d \Gamma_\o^2$, corresponding
to the wind around a Wolf-Rayet GRB progenitor that expelled $10^{-5}\, M_\odot {\rm yr^{-1}}$
at a terminal velocity of $10^3\, {\rm km\, s^{-1}}$. The result is
\begin{equation}
  t_d = 13\; \frac{z+1}{3} \frac{E_{53}}{A} \left(\frac{\Gamma_\o}{100}\right)^{-4} \; {\rm s} 
\end{equation}
Substituting in equation (\ref{t2}),
\begin{equation}
 \tau (t_d) = 6 \left( \frac{z+1}{3} \right)^5 \frac{A}{E_{53}} (F_{-6} \veps_8) \Gamma_\oo^{-2} 
\end{equation}
thus $\tau (t_d) = 1$ if $\Gamma_\o$ is
\begin{equation}
 \tildaG = 250 \left(\frac{z+1}{3}\right)^{5/2} \left(\frac{A F_{-6} \veps_8}{E_{53}}\right)^{1/2}
\end{equation}

 For $\Gamma_\o > \tildaG$, the $\tau$ falls below unity before $t_d$, at
\begin{equation}
 t_\pm \simeq 80 \left(\frac{z+1}{3}\right)^{6} (F_{-6} \veps_8) \left(\frac{\Gamma_\o}{100}\right)^{-6}\, {\rm s}
\end{equation}
At $t < t_\pm$, the emergent flux rises as $F = F_\o e^{-\tau} \propto \exp(-t_\pm/t)$, to a
plateau that begins after $t_\pm$ and lasts until $t_d$. The left panel of Figure 6 shows 
the attenuated light-curves obtained for $\Gamma_\o > \tildaG$. For the $\Gamma_\o$ shown, the plateau 
is not developed; a higher $\Gamma_\o$ would display that plateau, but at times $t < t_d \ll 1$ s and, 
hence, cannot be monitored. Thus, for $\Gamma_\o > \tildaG$, the attenuated LAT light-curve still 
displays a peak at $t_p \siml t_d$.

 For $\Gamma_\o < \tildaG$, the  optical thickness falls below unity after $t_d$, at a time $t_\pm$ that 
depends only on the dynamical parameter $E/A$ and the intrinsic afterglow flux (at some fixed time after $t_d$), 
as all information about $\Gamma_\o$ is lost after the deceleration time. At $t_d<t<t_\pm$, the emergent flux 
rises as $F = F_\o e^{-\tau} \propto t^{-1} \exp\{-(t_\pm/t)^{1/2}\}$, which is a slow rise to a peak at 
$t_p = t_\pm /4$. The right panel of Figure 6 shows the attenuated light-curves obtained for 
$\Gamma_\o < \tildaG$, displaying a broad hump after $t_d$. That hump cannot be significantly brighter 
than shown because a higher intrinsic flux is compensated by a larger optical thickness and the peak 
of the emergent light-curve gets even dimmer.

\section{Discussion}

 The results presented in the previous section indicate that pair-formation is negligible in
LAT afterglows if the ejecta initial Lorentz factor is higher than $\Gamma_\o \simeq 200$. 
In that case, the afterglow light-curve produced by the forward-shock energizing the ambient 
medium should be a broken power-law peaking at the ejecta deceleration time and without any 
evolution of the LAT spectrum.
At the other extreme, if $\Gamma_\o$ is too low (e.g. $\Gamma_\o = 50$), the afterglow
optical thickness $\tau$ to pair-formation could be too high and the attenuated flux too low to
be detected by LAT.

 For an intermediate range $50 \siml \Gamma_\o \siml 200$, photon-photon absorption is important
and modifies the broken power-law light-curve of the intrinsic (unatenuatted) forward-shock emission 
above 100 MeV, yielding any of the following:
\begin{enumerate}
\vspace*{-2mm}
 \item a light-curve peak prior to the onset of deceleration, with a spectral softening during that peak
\vspace*{-2mm}
 \item a post-peak flux power-law decay steeper than measured at lower photon energies
       (if the unabsorbed energy spectrum is harder then $F_\veps \propto \veps^{-4/3}$),
       accompanied by a spectral softening 
\vspace*{-2mm}
 \item a flux power-law decay slower than expected (if the intrinsic energy spectrum is 
       softer than $F_\veps \propto \veps^{-4/3}$), simultaneous with a spectral hardening
\vspace*{-2mm}
 \item a convex light-curve decay, accompanied by a spectral softening 
\vspace*{-2mm}
 \item a double peaked light-curve, for a soft $F_\veps \propto \veps^{-2}$ intrinsic spectrum,
       with the second peak accompanied by a spectral hardening
\vspace*{-2mm}
 \item a light-curve peak at the deceleration time and a spectral hardening for the entire afterglow
\vspace*{-2mm}
 \item a long-lived hump/plateau, accompanied by a spectral hardening.
\end{enumerate} 

 The first five features above are produced by the ejecta interaction with a homogeneous medium,
when the optical thickness to pair-formation $\tau$ increases with time before the deceleration 
time $t_d$ (the peak of the intrinsic afterglow light-curve), and increases/decreases after $t_d$ 
if the intrinsic afterglow spectrum is harder/softer than $F_\veps \propto \veps^{-4/3}$. 
The last two features listed above arise when the ambient medium has a wind-like particle radial 
distribution, in which case $\tau$ always decreases with time.
The behaviour of $\tau$ (increasing or decreasing) relates the corresponding afterglow feature to
a spectral evolution (softening or hardening, respectively). 

 {\sl Steep decays}.
 The afterglows 080916C (Abdo et al 2009b), 090510 (Ghirlanda et al 2010, Ackermann et al 2010a), 
090902B (Abdo et al 2009b), 090926A (Ackermann et al 2011), 100116, 100414, 110625 (Tam et al 2012), 
and 110731 (Ackermann et al 2013b) display a $F_{> 100 \, {\rm MeV}} \propto t^{-1.5}$ flux decay
or faster (Ghisellini et al 2010 attribute such fast decays to a highly radiative forward-shock).
XRT coverage overlapping with the LAT observations exists for 090510, 100414, 110625, and 110731, 
showing X-ray flux decays $F_{0.3-10 \; {\rm keV}} \propto t^{-0.7}$, $\propto t^{-0.6}$, 
$\propto t^{-0.6}$, and $\propto t^{-1.1}$, respectively, i.e. slower than at 100 MeV. 
Thus, the above LAT flux decays are faster than expected (from the X-rays) and the reason for 
those fast decays is not a radiative blast-wave.
The afterglow 100116 shows a marginally-significant softening by $\delta \beta = 0.9 \pm 0.9$, 
while 110625 has a more significant softening of $\delta \beta = 1.2 \pm 0.7$. 
The spectrum of afterglow 090926A hardens and, for the rest, no spectral evolution can be 
measured within the uncertainty of individual spectral slope measurements ($\sigma_{\beta} = 0.4$). 
For the last six LAT afterglows listed above, photons of more than 2 GeV were detected after $\sim 100$ s.
If such high-energy photons are associated with the afterglow at such late times, then photon-photon 
absorption plays no role in explaining those fast flux decays.

{\sl Convex light-curves}.
 The afterglow 080916C shows a convex light-curve (i.e. a steepening) but without a spectral softening.
A convex light-curve is compatible with the LAT flux upper limits that are below the extrapolation of the 
flux decay seen at earlier times for the afterglows 090217 (Ackermann et al 2010b) (which shows no 
clear spectral evolution), 100620 (which displays a spectral softening by $\delta \beta = 2.2 \pm 1.4$),
and 110120 (whose spectrum hardens by $\delta \beta = -0.8 \pm 0.7$, together with the detection
of a 2 GeV photon at 70 s, both suggesting that its emission is optically thin).
 As such light-curves are expected from a fast increase of $\tau$, perhaps caused by energy 
injection in the forward shock, their origin in this process should be corroborated with the 
corresponding plateau or slow decay in the X-ray light-curve. None of the above four LAT afterglows
have simultaneous XRT coverage.

{\sl Afterglow rises}.
 LAT light-curves with rises are seen for afterglows 080916C, 090510, 090902B, 090926A, and 110731,
with peaks at 1-10 s. 
A spectral evolution throughout the entire afterglow is evidenced only for 090926A,
with a spectral hardening from $\beta = 1.6 \pm 0.2$ at 5-10 s to $\beta = 0.7 \pm 0.2$ after 100 s. 
Furthermore, the LAT spectrum displays a cut-off at $\sim 2$ GeV and at 3-20 s, and photons above 
2 GeV appear after 20 s. Thus, this afterglow offers good evidence for the optical thickness to 
pair-formation decreasing throughout the afterglow. At late times, the spectrum should be unattenuated; 
being harder than $\beta = 4/3$, it implies that the decrease in $\tau$ should be associated with a
wind-like ambient medium. 

 Two-peaked light-curves and long-lived humps (or plateaus) could be difficult to observe with LAT, 
given that the expected flux at the second peak or during the plateau is strongly attenuated by
photon-photon absorption.

 An interesting effect of pair-formation is that, if the ambient medium is homogeneous, then there 
is an upper limit to the epoch of the light-curve peak that an adiabatic (no energy injection) forward 
shock can yield (eq \ref{tpmax}), that upper limit having a moderate dependence on the afterglow 
energy-to-ambient density ratio. Later occurring peaks could be produced by a wind-like medium. 
None of the LAT afterglows with light-curve peaks violate the upper limit for a homogeneous medium, 
nevertheless that does not identify the type of external medium because it may just be a selection effect:
later occurring peaks are intrinsically dimmer and harder to detect. Instead the identification of 
the ambient medium stratification can be done reliably from the evolution of the afterglow spectrum
(a softening of the afterglow spectrum is obtained only for a homogeneous medium). 

 In conclusion, there is tentative evidence that LAT may have observed the effects of photon-photon 
attenuation in two afterglows: 090926A (peaked ligh-curve and spectral hardening) and 100620 
(convex light-curve and spectral softening), and maybe also in 110625 (steep decay and spectral 
softening). 
Aside from the difficulty of measuring with LAT the spectral evolution that should accompany the 
above-listed light-curve features arising from photon-photon attenuation, the paucity of afterglows 
displaying the expected pair-formation features may be due to the narrow range of ejecta Lorentz 
factors $\Gamma_\o$ required for those features to occur ($\Gamma_\o \siml 200$) and for the emergent 
attenuated emission not to be too dim ($\Gamma_\o \simg 50$).
Conversely, the general lack of "self-absorption" features in LAT afterglow light-curves implies
that $\Gamma_o > 200$, which is consistent with the lower limits on the Lorentz factor inferred 
for GBM bursts (Ackermann et al 2013a).

\acknowledgments{This work was supported by an award from the Laboratory Directed Research and 
   Development program at the Los Alamos National Laboratory}

\end{document}